\documentstyle[twoside,epsfig]{article}
\newcommand{\agt}{\raise.3ex\hbox{$>$\kern-.75em\lower1ex\hbox{$\sim$}}~}
\newcommand{\alt}{\raise.3ex\hbox{$<$\kern-.75em\lower1ex\hbox{$\sim$}}~}

\catcode`\@=11
\long\def\@makefntext#1{
\protect\noindent \hbox to 3.2pt {\hskip-.9pt  
$^{{\eightrm\@thefnmark}}$\hfil}#1\hfill}		

\def\@makefnmark{\hbox to 0pt{$^{\@thefnmark}$\hss}}	
	
\def\ps@myheadings{\let\@mkboth\@gobbletwo
\def\@oddhead{\hbox{}
\rightmark\hfil\eightrm\thepage}   
\def\@oddfoot{}\def\@evenhead{\eightrm\thepage\hfil
\leftmark\hbox{}}\def\@evenfoot{}
\def\sectionmark##1{}\def\subsectionmark##1{}}

\oddsidemargin=\evensidemargin
\newcounter{sectionc}\newcounter{subsectionc}\newcounter{subsubsectionc}
\renewcommand{\section}[1] {\vspace{12pt}\addtocounter{sectionc}{1} 
\setcounter{subsectionc}{0}\setcounter{subsubsectionc}{0}\noindent 
	{\tenbf\thesectionc. #1}\par\vspace{5pt}}
\renewcommand{\subsection}[1] {\vspace{12pt}\addtocounter{subsectionc}{1} 
	\setcounter{subsubsectionc}{0}\noindent 
	{\bf\thesectionc.\thesubsectionc. {\kern1pt \bfit #1}}\par\vspace{5pt}}
\renewcommand{\subsubsection}[1] {\vspace{12pt}\addtocounter{subsubsectionc}{1}
	\noindent{\tenrm\thesectionc.\thesubsectionc.\thesubsubsectionc.
	{\kern1pt \tenit #1}}\par\vspace{5pt}}
\newcommand{\nonumsection}[1] {\vspace{12pt}\noindent{\tenbf #1}
	\par\vspace{5pt}}

\newcounter{appendixc}
\newcounter{subappendixc}[appendixc]
\newcounter{subsubappendixc}[subappendixc]
\renewcommand{\thesubappendixc}{\Alph{appendixc}.\arabic{subappendixc}}
\renewcommand{\thesubsubappendixc}
	{\Alph{appendixc}.\arabic{subappendixc}.\arabic{subsubappendixc}}

\renewcommand{\appendix}[1] {\vspace{12pt}
        \refstepcounter{appendixc}
        \setcounter{figure}{0}
        \setcounter{table}{0}
        \setcounter{lemma}{0}
        \setcounter{theorem}{0}
        \setcounter{corollary}{0}
        \setcounter{definition}{0}
        \setcounter{equation}{0}
        \renewcommand{\thefigure}{\Alph{appendixc}.\arabic{figure}}
        \renewcommand{\thetable}{\Alph{appendixc}.\arabic{table}}
        \renewcommand{\theappendixc}{\Alph{appendixc}}
        \renewcommand{\thelemma}{\Alph{appendixc}.\arabic{lemma}}
        \renewcommand{\thetheorem}{\Alph{appendixc}.\arabic{theorem}}
        \renewcommand{\thedefinition}{\Alph{appendixc}.\arabic{definition}}
        \renewcommand{\thecorollary}{\Alph{appendixc}.\arabic{corollary}}
        \renewcommand{\theequation}{\Alph{appendixc}.\arabic{equation}}
        \noindent{\tenbf Appendix \theappendixc #1}\par\vspace{5pt}}
\newcommand{\subappendix}[1] {\vspace{12pt}
        \refstepcounter{subappendixc}
        \noindent{\bf Appendix \thesubappendixc. {\kern1pt \bfit #1}}
	\par\vspace{5pt}}
\newcommand{\subsubappendix}[1] {\vspace{12pt}
        \refstepcounter{subsubappendixc}
        \noindent{\rm Appendix \thesubsubappendixc. {\kern1pt \tenit #1}}
	\par\vspace{5pt}}

\newcommand{\textlineskip}{\baselineskip=13pt}
\newcommand{\smalllineskip}{\baselineskip=10pt}

\def\eightcirc{
\begin{picture}(0,0)
\put(4.4,1.8){\circle{6.5}}
\end{picture}}
\def\eightcopyright{\eightcirc\kern2.7pt\hbox{\eightrm c}}

\def\abstracts#1#2#3{{
	\centering{\begin{minipage}{4.5in}\footnotesize\baselineskip=10pt
	\parindent=0pt #1\par 
	\parindent=15pt #2\par
	\parindent=15pt #3
	\end{minipage}}\par}} 

\renewenvironment{thebibliography}[1]
	{\frenchspacing
	 \ninerm\baselineskip=11pt
	 \begin{list}{\arabic{enumi}.}
	{\usecounter{enumi}\setlength{\parsep}{0pt}
	 \setlength{\leftmargin 12.7pt}{\rightmargin 0pt} 
	 \setlength{\itemsep}{0pt} \settowidth
	{\labelwidth}{#1.}\sloppy}}{\end{list}}

\newcounter{itemlistc}
\newcounter{romanlistc}
\newcounter{alphlistc}
\newcounter{arabiclistc}

\newcommand{\fcaption}[1]{
        \refstepcounter{figure}
        \setbox\@tempboxa = \hbox{\footnotesize Fig.~\thefigure. #1}
        \ifdim \wd\@tempboxa > 5in
           {\begin{center}
        \parbox{5in}{\footnotesize\smalllineskip Fig.~\thefigure. #1}
            \end{center}}
        \else
             {\begin{center}
             {\footnotesize Fig.~\thefigure. #1}
              \end{center}}
        \fi}

\newcommand{\tcaption}[1]{
        \refstepcounter{table}
        \setbox\@tempboxa = \hbox{\footnotesize Table~\thetable. #1}
        \ifdim \wd\@tempboxa > 5in
           {\begin{center}
        \parbox{5in}{\footnotesize\smalllineskip Table~\thetable. #1}
            \end{center}}
        \else
             {\begin{center}
             {\footnotesize Table~\thetable. #1}
              \end{center}}
        \fi}

\def\@citex[#1]#2{\if@filesw\immediate\write\@auxout
	{\string\citation{#2}}\fi
\def\@citea{}\@cite{\@for\@citeb:=#2\do
	{\@citea\def\@citea{,}\@ifundefined
	{b@\@citeb}{{\bf ?}\@warning
	{Citation `\@citeb' on page \thepage \space undefined}}
	{\csname b@\@citeb\endcsname}}}{#1}}

\newif\if@cghi
\def\cite{\@cghitrue\@ifnextchar [{\@tempswatrue
	\@citex}{\@tempswafalse\@citex[]}}
\def\citelow{\@cghifalse\@ifnextchar [{\@tempswatrue
	\@citex}{\@tempswafalse\@citex[]}}
\def\@cite#1#2{{$\null^{#1}$\if@tempswa\typeout
	{IJCGA warning: optional citation argument 
	ignored: `#2'} \fi}}

\def\pmb#1{\setbox0=\hbox{#1}
	\kern-.025em\copy0\kern-\wd0
	\kern.05em\copy0\kern-\wd0
	\kern-.025em\raise.0433em\box0}

\def\fnt#1#2{\footnotetext{\kern-.3em
	{$^{\mbox{\scriptsize #1}}$}{#2}}}

\def\@makefnmark{\hbox to 0pt{$^{\@thefnmark}$\hss}}	
	
\def\ps@myheadings{%
    \let\@oddfoot\@empty\let\@evenfoot\@empty
    \def\@evenhead{\slshape\leftmark\hfil}
    \def\@oddhead{\hfil{\slshape\rightmark}}
    \let\@mkboth\@gobbletwo
    \let\sectionmark\@gobble
    \let\subsectionmark\@gobble
    }
\font\tenrm=cmr10
\font\tenit=cmti10 
\font\tenbf=cmbx10
\font\bfit=cmbxti10 at 10pt
\font\ninerm=cmr9

\font\eightrm=cmr8

\def\qed{\hbox{${\vcenter{\vbox{			
   \hrule height 0.4pt\hbox{\vrule width 0.4pt height 6pt
   \kern5pt\vrule width 0.4pt}\hrule height 0.4pt}}}$}}

\newcommand{\be}{\begin{equation}}
\newcommand{\ee}{\end{equation}}
\newcommand{\ba}{\begin{eqnarray}}
\newcommand{\ea}{\end{eqnarray}}

\begin{document}
\setlength{\textheight}{7.7truein}  

\markboth{\protect{\footnotesize\it I. A. Shovkovy}}
{\protect{\footnotesize\it Collective modes in color 
superconducting matter}}

\normalsize\textlineskip

\vspace*{0.88truein}

\centerline{\bf COLLECTIVE MODES IN COLOR SUPERCONDUCTING
MATTER\footnote{This work was supported by the U.S.
        Department of Energy Grant No.~DE-FG02-87ER40328.}}
\vspace*{0.37truein}
\centerline{\footnotesize IGOR A. SHOVKOVY\footnote{On leave of absence from
                    Bogolyubov Institute for Theoretical
                    Physics, 03143, Kiev, Ukraine.}}
\baselineskip=12pt
\centerline{\footnotesize\it School of Physics and Astronomy,
University of Minnesota, 116 Church Street S.E.}
\baselineskip=10pt
\centerline{\footnotesize\it Minneapolis, MN 55455, USA}

\vspace*{0.21truein}
\abstracts{The properties of plasmons, Nambu-Goldstone bosons and gapless
Carlson-Goldman collective modes in color-flavor locked phase of color
superconducting dense quark matter at finite temperature are reviewed. A
possibility of a kaon condensation with an abnormal number of the NG
bosons is also discussed.}{}{}

\vspace*{1pt}\textlineskip      
\section{Introduction}          
\vspace*{-0.5pt}
\noindent

In this talk I review the approach and results of recent studies\cite{GS}
on collective modes in the color-flavor locked (CFL) phase of dense quark
matter at finite temperature. Also, I am going to mention about an
interesting possibility of a kaon condensation appearing on top of the CFL
phase.\cite{BS,KR,MS,SSSTV}

Let me start by noticing that there is a growing belief in the literature
that the cores of some compact stars might be made of a color
superconducting quark matter. Although this was not proved in any reliable
way, such a possibility at least seems to be likely. Indeed, different
theoretical studies suggest that the value of the superconducting order
parameter could be as large as $10$ to $100$ MeV at densities just a few
times larger than the density of the ordinary nuclear
matter.\cite{WS1,PR1,Son,us} If this is really so, the observed properties
of compact stars might show up an indication of a color superconducting
state.\cite{stars,supernova}

The CFL phase of dense quark matter has very interesting properties. Most
of them have already been recognized in the original paper.\cite{CFL}
Thus, I am not going to describe them here. For the purposes of this
presentation, it would suffice to mention only a few key facts that play a
significant role in the analysis.

First of all, it would be appropriate to recall the color-flavor structure
of the order parameter. In general, it contains an antitriplet-antitriplet
$(\bar{3},\bar{3})$ and a sextet-sextet $(6,6)$ contribution:\cite{PR1}
\be
\Delta^{ab}_{ij} = \Delta_{\bar{3},\bar{3}} 
(\delta^{a}_{i} \delta^{b}_{j}- \delta^{a}_{j} \delta^{b}_{i})
+\Delta_{6,6} 
(\delta^{a}_{i} \delta^{b}_{j}+\delta^{a}_{j} \delta^{b}_{i}) ,
\ee
where the flavor ($i,j=1,2,3$) and color ($a,b=1,2,3$) indices are
explicitly displayed. The dominant contribution to the order parameter is
$\Delta_{\bar{3},\bar{3}}$, but a nonzero although small admixture of
$\Delta_{6,6}$ always appears.\cite{CFL,us2,Spt}

In the color superconducting phase, there are two types of quark
quasiparticles which form octet and singlet representations under the
vacuum symmetry group $SU(3)_{c+L+R}$. Their one-particle spectra develop
different energy gaps around the Fermi surface. The values of the gaps are
$\Delta_{1}=2(\Delta_{\bar{3},\bar{3}}+2\Delta_{6,6})$ and 
$\Delta_{8}=-(\Delta_{\bar{3},\bar{3}}-\Delta_{6,6})$, respectively. In
the case of a pure $(\bar{3},\bar{3})$ order parameter, these gaps are
related as follows: $\Delta_{1} = -2\Delta_{8}$. Without loosing
generality (unless, of course, $\Delta_{1}= \pm \Delta_{8}$), I consider
this relation to hold approximately at both zero and nonzero temperatures,
i.e., $\Delta_{1} \approx -2\Delta_{8}=-2|\Delta_{T}|$.

\section{Strategy}
\noindent
The aim of this study is to reveal the whole class of collective modes in
the CFL phase that couple to either vector or axial color
currents.\cite{GS} This ambitious task is solved (at large densities where
the color interaction becomes weak) by studying the locations of poles in
the corresponding current-current correlation functions,
\ba
\int d^{4} x e^{iq x} \langle T
j^{\mu,A}{\tiny (x)} j^{\nu,B}{\tiny (0)} \rangle
\equiv \langle j^{\nu,B} j^{\mu,A} \rangle_{q}
\sim \sum_{n} \frac{(\mbox{\ldots})^{\mu\nu,AB}}
{q_{0}^{2}-\Pi^{(n)}(q)},\\
\int d^{4} x e^{iq x} \langle T
j^{\mu,A}_{5}{\tiny (x)} j^{\nu,B}_{5}{\tiny (0)} \rangle
\equiv \langle j^{\nu,B}_{5} j^{\mu,A}_{5} \rangle_{q}
\sim \sum_{n} \frac{(\ldots)^{\mu\nu,AB}}
{q_{0}^{2}-\Pi^{(n)}_{5}(q)}, 
\ea
where the sum runs over all poles. In order to derive such correlation
functions it is sufficient to know only their irreducible parts, i.e.,
the polarization tensors.

In the broken phase, the polarization tensor (related to the vector color
current) contains two qualitatively different contributions. One of them
is the ordinary one-loop quark contribution, while the other comes from
the ``would be Nambu-Goldstone (NG) bosons". This latter contribution is
subtle. It appears in all nonunitary gauges as a result of proper handling
the gauge invariance of the model.\cite{bs-us} A similar contribution also
appears naturally in the effective theory.\cite{CD} Diagrammatically, the
full expression is presented as follows:\\
\epsfxsize=9.5cm
\epsfbox[75 260 400 350]{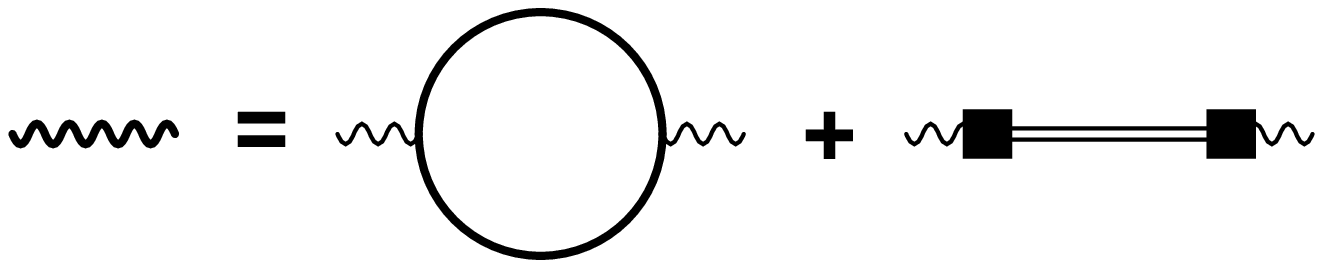}\\
where the double line denotes the would be NG bosons.  The corresponding
analytical expression was derived based on arguments of gauge symmetry in
our recent paper.\cite{GS} It reads
\be
\Pi^{(full)}_{\mu\nu}(q)
=\Pi_{1}(q) O^{(1)}_{\mu\nu}(q) + \left(\Pi_{2}(q)
+\frac{[\Pi_{4}(q)]^{2}}{\Pi_{3}(q)} \right)O^{(2)}_{\mu\nu}(q),
\label{pol-ten}
\ee
where the component functions $\Pi_{n}(q)$ are extracted from 
the expansion of the one-loop contribution of quarks to the 
polarization tensor,
\be 
\Pi_{\mu\nu}(q) =
 \Pi_{1}(q)  O^{(1)}_{\mu\nu}(q)
+\Pi_{2}(q)  O^{(2)}_{\mu\nu}(q)
+\Pi_{3}(q)  O^{(3)}_{\mu\nu}(q)
+\Pi_{4}(q)  O^{(4)}_{\mu\nu}(q) .
\label{Pi-gen}
\ee
The tensors $O^{(n)}_{\mu\nu}(q)$ (two of which are not transverse) 
are defined as follows:
\ba
O^{(1)}_{\mu\nu}(q) = g_{\mu\nu}-u_{\mu} u_{\nu}
+\frac{\vec{q}_{\mu}\vec{q}_{\nu}}{|\vec{q}|^{2}},
& \quad &
O^{(2)}_{\mu\nu}(q) = u_{\mu} u_{\nu}
-\frac{\vec{q}_{\mu}\vec{q}_{\nu}}{|\vec{q}|^{2}}
-\frac{q_{\mu}q_{\nu}}{q^{2}},
\label{def-O12} \\
O^{(3)}_{\mu\nu}(q) = \frac{q_{\mu}q_{\nu}}{q^{2}},
& \quad &
O^{(4)}_{\mu \nu}(q) =
O^{(2)}_{\mu \lambda} u^{\lambda} \frac{q_{\nu}}{|\vec{q}|}
+\frac{q_{\mu}}{|\vec{q}|}u^{\lambda} O^{(2)}_{\lambda \nu}.
\label{def-O34}
\ea
with $u_{\mu} = (1,0,0,0)$ and $\vec{q}_{\mu} = q_{\mu} - (u\cdot q)  
u_{\mu}$. It is remarkable to notice that the contribution of the would be
NG bosons [i.e., $\Pi_{4}^{2}/\Pi_{3}$ term in Eq.~(\ref{pol-ten})] is
defined in terms of the nontransverse components of the quark contribution
itself.

The polarization tensor in Eq.~(\ref{pol-ten}) determines the 
(connected) correlation function
\be
\langle j^{A}_{\mu} j^{B}_{\nu} \rangle_{q}
= \delta^{AB} \left[
\frac{ q^{2} \Pi_{1}}{q^{2}+\Pi_{1}} O^{(1)}_{\mu\nu}(q) 
+\frac{ q^{2} \left[\Pi_{2} \Pi_{3} + (\Pi_{4})^{2}\right]}
{(q^{2}+ \Pi_{2}) \Pi_{3} + (\Pi_{4})^{2}}
O^{(2)}_{\mu\nu}(q)
\right]. 
\label{cur-cur}
\ee
The locations of poles in this function define the dispersion 
relations (as well as screening effects) of the magnetic and 
electric type collective excitations,
\ba 
q^{2} + \Pi_{1}(q) &=& 0, 
\quad \mbox{``magnetic modes"},
\label{spectrum-mag}\\
\left[q^{2} + \Pi_{2}(q)\right] \Pi_{3}(q) + [\Pi_{4}(q)]^{2} 
&=& 0, 
\quad \mbox{``electric modes"}. 
\label{spectrum-el}
\ea
In a similar way, one derives the dispersion relations of the 
pseudoscalars coupled to the axial color currents, $\Pi_{3}(q) =0$.
These pseudoscalars are nothing else but the NG bosons associated 
with breaking of the global chiral symmetry.

\section{Plasmons and ``light" plasmons} 
\noindent 
Dense quark matter is a non-Abelian plasma. Similarly to an ordinary
plasma, the non-Abelian plasma supports collective oscillations of the
(color) charge density. By analogy, it is natural to call the
corresponding modes plasmons.

Plasmons exist in the normal as well as in the superconducting phase.  
The gap in their spectrum, usually called plasma frequency ($\omega_{p}$), 
is directly
related to the quark density in the system. Thus, by considering the limit
$q_{0} \gg |\Delta_{T}|$ (large energy) and $|\vec{q}|\to 0$ (long
wavelength) in Eqs.~(\ref{spectrum-mag}) and (\ref{spectrum-el}), we
derive the dispersion relations for both the magnetic and electric modes
\be
q_{0}^{2} = \omega_{p}^{2}, \quad \mbox{with} \quad
\omega_{p}\equiv \frac{g \mu }{\sqrt{2}\pi} \gg |\Delta_{T}|.
\label{Omega=omega_p}
\ee
As in the case of ordinary metals, the plasmon frequency is essentially
unaffected by a small (compared to the chemical potential $\mu$)  
temperature of the system and by the presence of the superconducting gap.
This should have been expected.

Much more interesting are the ``light" plasmons that were also discovered
in the CFL phase of dense quark matter.\cite{GS} These light plasmon modes
have no analogues in ordinary metals.  Their appearance in the CFL phase
seems to be directly related to the presence of two different types of
quark quasiparticles having nonequal gaps in their spectra, $|\Delta_{1}|
\neq |\Delta_{8}|$.

In the long wavelength limit, the energy of the light plasmon is always
smaller than the threshold of producing a pair of quasiparticles, $|q_{0}|
< 2 |\Delta_{0}|$. At zero temperature, in particular, one derives $q_0
=m_{\Delta} \approx 1.362 |\Delta_{0}|$. The numerical calculation of the
temperature dependence of the light plasmon mass (i.e., the gap in its 
spectrum) $m_{\Delta} (T)$ is also
available.\cite{GS} Qualitatively, $m_{\Delta} (T)$ is a monotonically
decreasing function of temperature which approaches zero as $T\to T_{c}$.
Besides that, in the whole range of temperatures $0<T<T_{c}$, it remains
larger than $1.362 |\Delta_{T}|$ and smaller than the threshold
$2|\Delta_{T}|$. It is clear, therefore, that light plasmon excitations
are stable with respect to decays into quark quasiparticles. It should be
mentioned, however, that other decay channels (for example, involving the
NG bosons) might eventually lead to a nonzero width. Unfortunately, this
question has not been studied yet in detail.

Before concluding this section, I should note that an indication of a
light plasmon could also be seen in the derivative expansion.\cite{CGN}
However, the derivative expansion approach (treating the ratios
$|q_0|/|\Delta_{T}|$ and $|\vec{q}|/|\Delta_{T}|$ as small parameters)
overestimates the value of the mass.

\section{Nambu-Goldstone bosons} 
\noindent 
The global chiral symmetry is broken in the CFL phase of dense QCD. This
means that an octet of pseudoscalar NG bosons should appear in the
spectrum. The dispersion relation of the NG bosons reads: $\Pi_{3}(q)=0 $.
This relation could be analytically studied at small temperatures and in
the nearcritical region.

In the first case ($T\ll |\Delta_{0}|$), the dispersion relation 
of the NG bosons reads\cite{GS}
\ba
q_0=\frac{|{\vec q}|}{\sqrt3}
\left[1-i\frac{5\sqrt{2}\pi}{4(21-8\log2)}
e^{-\sqrt{\frac{3}{2}}\frac{|\Delta_{0}|}{T}}\right].
\label{E13}
\ea
Note that this relation reveals an exponentially small imaginary part when
$T\to 0$. It is worthwhile to mention, however, that the width of the NG
bosons might also get corrections due to the interactions of the NG bosons
with one another. Since their decay constant is of the same order as the
chemical potential $\mu$, such corrections are expected to be
parametrically suppressed by some power of the ratio $T/\mu$.\footnote{A
simple estimate shows that the power low corrections to the imaginary part
in Eq.~(\ref{E13}) are negligible for $T \agt 0.01 g |\Delta_{0}|$. This
leaves a window of temperatures, $0.01 g |\Delta_{0}| \ll T \ll
|\Delta_{0}|$, where the dispersion relation (\ref{E13}) is qualitatively
correct.}

When $T\to T_{c}-0$, the dispersion relation reads
\be
q_{0}
=\frac{|\Delta_{T}|}{T}(\pm x_{ng}^{*} - i y_{ng}^{*})
|\vec{q}|, \label{result-NG}
\ee
where $x_{ng}^{*}\approx 0.215$ and $y_{ng}^{*}\approx 0.245$.  This
result shows that the NG bosons have a large width. This is hardly a
surprise in the nearcritical region where the gap in the quasiparticle
spectrum is vanishingly small. I would like to point out, however, that
the numerical analysis reveals well pronounced peaks in the NG boson
spectral density and the peak locations scale with momentum in
accordance with a linear dispersion law.\cite{GS}

\section{Carlson-Goldman collective modes}
\noindent
The so-called Carlson-Goldman (CG) gapless modes were experimentally
discovered by Carlson and Goldman about a three decades ago.\cite{CG} One
of the most intriguing interpretation connects such modes with a revival
of the NG bosons in the superconducting phase where the Anderson-Higgs
mechanism should commonly take place.  In the two fluid description, the
CG modes are seen as oscillations of the superfluid and the normal
components in opposite directions.\cite{SchSch} The local charge density
remains zero in such oscillations, providing favorable conditions for
gapless modes.

The CG modes can only appear in the vicinity of the critical temperature
(in the broken phase), where a large number of thermally excited
quasiparticles leads to partial screening of the Coulomb interaction, and
the Anderson-Higgs mechanism becomes inefficient. As a result, such modes
could exist only in a finite (possibly very small) vicinity of the
critical temperature. Because of the Landau damping induced by the
quasiparticles, however, the CG modes could easily become overdamped in
the nearcritical region.\cite{Artem79,Tak97} In order to make them
observable, one should use dirty systems in which scattering of
quasiparticles on impurities tends to reduce the damaging effect of
Landau damping.

In the nearcritical region, the dispersion relation of the CG collective
modes in the CFL phase could be determined analytically. It is remarkable
that the result closely resembles the dispersion relation of the NG bosons
in Eq.~(\ref{result-NG}),
\be
q_{0} =\frac{|\Delta_{T}|}{T}(\pm x_{cg}^{*} - i y_{cg}^{*})
|\vec{q}|,
\label{result}
\ee
where $x_{cg}^{*} \approx 0.193$ and $y_{cg}^{*}\approx 0.316$. As in
the case of the NG bosons, the width of the CG modes is quite large. It
should be emphasized here that the above result was obtained in the
clean (no impurities) limit of dense quark matter. Notice that some 
impurities might come to game if the phase transition in temparature 
is a weak first order one. In this case, the bubbles of the new phase
may play the role of impurities. Of course, this assumes that the 
transition is sufficiently weak, so that the value of the gap would 
have a chance to get quite small on the broken side (say, less than
about $0.05 T_{c}$).

As in the case of the NG bosons, the numerical analysis of the electric
gluon spectral density reveals distinguishable (although less pronounced
than in the case of the NG bosons) maxima whose locations scale with
momenta in accordance with a linear dispersion law.\cite{GS}

\section{Nambu-Goldstone bosons in phase with kaon condensation}
\noindent
While the CFL phase is the ground state of the three flavor dense quark
matter in the chiral limit,\cite{CFL} this may not be the case when the
quarks have nonzero masses. By making use of an auxiliary gauge symmetry,
it was recently suggested\cite{BS} that the originally derived low-energy
action\cite{CasGat,SonSt} should be modified. In Minkowski space, the
corresponding Lagrangian density should read\cite{BS}
\ba
{\cal L}_{eff} &=& \frac{f_\pi^2}{4} {\rm Tr}\left[
 \nabla_0\Sigma\nabla_0\Sigma^\dagger - v_\pi^2
 \partial_i\Sigma\partial_i\Sigma^\dagger \right]
+\frac{1}{2} \left[ (\partial_0 \eta^{\prime})^{2}
- v_{\eta^{\prime}}^2  (\partial_i\eta^{\prime})^{2} \right]
\nonumber \\
&&+ 2 c\left[\det(M) \mbox{Tr} \left( M^{-1} \Sigma
e^{\sqrt{\frac{2}{3}} \frac{i}{f_{\eta^{\prime}}} {\cal I}
\eta^{\prime}}\right) + h.c.\right],
\label{L-eff} 
\ea
where, by definition, ${\cal I}$ is a unit matrix in the flavor space, and
$\Sigma$ is a $3\times 3$ unitary matrix field which describes an octet of
the NG bosons. In the action, we took into account the $\eta^{\prime}$
field which couples to the NG octet when the quark masses are nonzero. At
the same time, the NG boson, related to breaking the baryon number, was
omitted in Eq.~(\ref{L-eff}). Its dynamics is not affected much by the
quark masses. The covariant time derivative in Eq.~(\ref{L-eff}) is
defined in terms of the quark mass matrix as follows:
\be
\nabla_0\Sigma = \partial_0 \Sigma
 + i \frac{M M^\dagger}{2p_F} \Sigma
 - i \Sigma \frac{ M^\dagger M}{2p_F}, \quad
\mbox{with} \quad
M=\mbox{diag}(m_{u},m_{d},m_{s}). 
\label{cov-der}
\ee
As is clear from this definition, the quark masses produce effective
chemical potentials for different flavors. In the case of a realistic
hierarchy of the quark masses with $m_{s} \gg m_{u} \simeq m_{d}$, these
chemical potentials may trigger a rearrangement of the CFL phase. In
particular, when the value of strange quark mass is large enough so that
the condition $m_{s}\agt (\Delta^{2}m_{u})^{1/3}$ is satisfied, the CFL
vacuum becomes unstable with respect to a kaon condensation.\cite{BS} The
new ground state is determined by a ``rotated" vacuum expectation value of
the $\Sigma$ field,
\be
\Sigma_{\alpha} \equiv \exp\left(i\alpha\lambda^{6}\right)
\exp\left(i\pi_{A}\lambda^{A}\right)
\simeq  \exp\left(i\alpha\lambda^{6}\right)\left(
1 + \frac{i\pi_{A}\lambda^{A}}{f_\pi}
- \frac{\pi_{A}\pi_{B}\lambda^{A}\lambda^{B}}{2f_\pi^2}
+ \ldots \right).
\label{new-gr-state}
\ee
The value of the angle $\alpha$ is determined from the condition that the
vacuum expectation value in Eq.~(\ref{new-gr-state}) corresponds to a
global minimum of the potential energy,
\be
\cos\alpha=\frac{4cp_{F}^{2}m_{u}(m_{s}+m_{d})}
{f_{\pi}^{2}(m_{s}^{2}-m_{d}^{2})^{2}}<1,
\label{alpha}
\ee
where\cite{SonSt}
\be
c = \frac{3 \Delta^{2}}{2\pi^{2}}
\quad \mbox{and} \quad
f_{\pi}^{2} = \frac{21-8\ln2}{36} \frac{\mu^{2}}{\pi^{2}}.
\ee
When the isospin symmetry is exact (i.e., $m_{u}=m_{d}$), the kaon
condensation breaks the $SU(2)\times U(1)_{Y}$ symmetry of the effective
action (\ref{L-eff}) down to $U(1)^{\prime}$.  This symmetry breaking
pattern (with {\em three} broken generators of a global symmetry) might
suggest that {\em three} NG bosons should appear in the low energy
spectrum of the system. The direct calculation shows, however, that only
{\em two} NG bosons appear.\cite{MS,SSSTV} In order to understand this
seeming paradox better, we consider a toy model that closely mimics the
dynamics of the kaon condensation. The Lagrangian density of the toy model
reads
\ba
{\cal L} &=& (\partial_{0} +i \mu )\Phi^{\dagger}
(\partial_{0} -i \mu )\Phi-v^2\partial_{i}\Phi^{\dagger}
\partial_{i}\Phi-m^{2}\Phi^{\dagger} \Phi
-\lambda(\Phi^{\dagger} \Phi)^{2},
\label{L-model}
\ea
where $\Phi^{T}=\frac{1}{\sqrt{2}} (\phi_{1}+i\phi_{2},\tilde{\phi}_{1}+
i\tilde{\phi}_{2})$ is a complex doublet field, replacing four kaon
degrees of freedom, and $v$ is a velocity parameter ($v < 1$). The
parameter $\mu$ corresponds to the effective chemical potential of
strangeness in the realistic model.

It is straightforward to derive the spectrum of excitations in the toy
model (\ref{L-model}) in the normal phase which is realized when $\mu<m$.
They are
\ba
\omega_{K_{0}}(q) = \omega_{K^{+}}(q) = \sqrt{m^{2}+v^2 q^2} -\mu ,
\label{m-mu} \\
\omega_{\bar{K}_{0}}(q) = \omega_{K^{-}}(q) = \sqrt{m^{2}+v^2 q^2}+\mu ,
\label{m+mu}
\ea
where we explicitly identified the four degrees of freedom of $\Phi$ with
the kaons. 

When the chemical potential becomes larger than the mass parameter,
$\mu > m$, the vacuum configuration with $\langle \Phi \rangle =0$ is no
longer a local minimum of the potential energy. This clearly indicates
that the system develops an instability with respect to forming a
condensate. In the true ground state, the field $\Phi$ has a nonzero
vacuum expectation value which, up to a global transformation, is $\langle
\Phi \rangle = (0, \phi_{0})^{T}$ and
\be
\phi_{0}^{2} = \frac{\mu^{2}-m^{2}}{2\lambda}.
\ee
This condensate breaks the initial $SU(2)\times U(1)$ symmetry group of
the toy model (\ref{L-model}) down to $U(1)^{\prime}$. It is interesting
to study the spectrum of excitations in this broken phase. The four
degrees of freedom of the complex doublet $\Phi$ now have the following
dispersion relations:
\ba
\omega_{1,2} =\sqrt{\mu^{2}+v^{2}q^{2}}\pm \mu,
\label{dsp1}
\ea
and 
\be
\tilde{\omega}_{1,2} =\sqrt{3\mu^{2}-m^{2} +v^{2}q^{2}
\pm \sqrt{(3\mu^{2}-m^{2})^{2} + 4\mu^{2}v^{2}q^{2} }}.
\label{dsp2}
\ee
Only one of the relations in Eq.~(\ref{dsp1}) and only one
of the relations in Eq.~(\ref{dsp2}) correspond to gapless excitations.  
In the far infrared region, the corresponding relations take the following
forms:
\be
\omega_{2} \simeq  \frac{v^{2}q^{2}}{2\mu}, \quad \mbox{and} \quad 
\tilde{\omega}_{2} \simeq \sqrt{\frac{\mu^{2}-m^{2}}{3\mu^{2}-m^{2}}}v q.
\label{NG}
\ee
Thus, there are only {\em two} gapless NG bosons in the low energy
spectrum of the model. The other two excitations have gaps $m_{1}=2\mu$
and $\tilde{m}_{1} = \sqrt{2(3\mu^{2}-m^{2})}$. Here one should notice
that the existence of only two gapless states should have been
expected based on a simple argument of continuity of the spectrum at
the critical point $\mu=m$. By approaching this point from the side of the
normal phase, we see from Eqs.~(\ref{m-mu}) and (\ref{m+mu}) that only
{\em two} out of the total four states become gapless as $\mu\to m-0$.

To avoid a possible confusion, it is instructive to mention that the
effective potential has {\em three} flat directions in the vicinity of the
vacuum configuration, as it should in the case of the symmetry breaking
$SU(2)\times U(1) \to U(1)^{\prime}$. It is the first order derivative
terms in the kinetic part of the action that prevent the appearance of the
third gapless mode. The same first order derivative terms break the
Lorentz invariance (and, maybe even more important, the discrete $C$, $CP$
and $CPT$ symmetries) of the model and the strong version of the Goldstone
theorem\cite{GSW} cannot be applied. In non-relativistic physics, the
known examples of systems with an abnormal number of the NG bosons are
ferromagnets\cite{fer} and the superfluid $^{3}$He in the so-called
A-phase.\cite{Volovik}

As is seen from Eq.~(\ref{NG}), the two NG bosons have qualitatively
different dispersion relations. One of them has a quadratic dispersion
relation, while the other has a linear dispersion relation. The existence
of the NG boson with a quadratic relation has an immediate implication ---
it prevents superfluidity in the system.\cite{MS} Indeed, according to the
Landau criterion,\cite{Lifshitz-Pitaevskii} the superfluidity is possible
only for flow velocities $v < v_c$, where $v_c = \mbox{min}_{i,q}
(\omega_{i}(q)/q)$ is the minimum taken over all excitation branches and
all values of momentum. The presence of even a single branch with $\omega
\sim q^2$ implies that $v_c = 0$ and, therefore, that there is no
superfluidity.

The toy model (\ref{L-model}) illustrates a general phenomenon of
spontaneous breaking of a global symmetry ($G\to H$) in systems with a
broken Lorentz symmetry. In particular, the number of NG bosons can be
smaller than the number of the generators $N_{G/H}$ in the coset space
$G/H$. In the case of dense quark matter with the kaon condensation, in
particular, half of the charged candidates to NG bosons have gaps in
their energy spectra as a result of 
spliting produced by an effective chemical potential. The
neutral NG boson candidates do not feel the chemical potential and, thus,
remain gapless. This observation allows us to propose the following
counting rule: the number of the gapless NG bosons equals to
$N_{NG}=N_{G/H}-N_{ch}$, where $N_{ch}$ is the number of charged
particle-antiparticle pairs among the NG boson candidates. In model
(\ref{L-model}), $N_{G/H}=3$ and $N_{ch}=1$, so that $N_{NG}=2$.

Because the isospin symmetry is not exact in the real world ($m_{u}\neq
m_{d}$), the effective action (\ref{L-eff}) has only $U(1)\times U(1)_{Y}$
symmetry. The kaon condensate breaks this symmetry down to
$U(1)^{\prime}$. As a result, only one NG boson (with a linear dispersion
relation) appears in the low energy spectrum. The corresponding phase of
matter is expected to be superfluid since there are no excitations with
the quadratic dispersion relations. The approximate isospin symmetry could
reveal itself only through the appearance of a light excitation, replacing
the NG boson with the quadratic relation, see Eq.~(\ref{NG}). When the
difference between up and down quark masses is very small, this light
excitation could considerably reduce the critical value of the superfluid
velocity.

\section{Conclusion and Outlook}
\noindent
The study of a large class of collective modes, reviewed in this talk,
gives a rather complete description of physical degrees of freedom in
the CFL phase of dense quark matter at zero and finite temperatures.
Some of the modes (light plasmons and the gapless CG modes)  have
very interesting properties and have no analogues in other phases.
Potentially this may play an important role in producing a key signal of
a color superconducting phase in nature. With this in view, it would be
of great interest to see, for example, whether the existence of the
gapless scalar CG modes could affect any thermodynamical or transport
properties of dense quark matter. And, in its turn, whether this could
have any profound effect on the evolution of forming compact stars.

In the case of nonzero quark masses, it is possible that the CFL phase is
modified due to the appearance of the kaon condensate.  The general
properties of collective excitations in such an exotic phase were also
briefly addressed in this presentation. This new state of dense matter
gives an interesting example of a system in which the number of the NG
bosons should not necessarily be equal to the number of the broken
generators. Related to this is the possibility of excitations with
(nearly) quadratic dispersion relations. The latter, if appear, may have
interesting observable implications. Indeed, as is known from statistical
physics, the thermodynamical quantities such as specific heat are quite
sensitive to the details of the low energy spectrum.

\nonumsection{Acknowledgements}

I would like to thank the organizers of the conference for giving me an
opportunity to present these results. Also, I would like to thank my
collaborators V.P. Gusynin and V.A. Miransky for interesting discussions
and valuable comments. Finally, I thank A.F.~Volkov and G.E.~Volovik for
bringing Refs.~[20] and [26] to my attention. This work was supported by
the U.S. Department of Energy Grant No.~DE-FG02-87ER40328.

\nonumsection{References}

\end{document}